\documentclass[useAMS,usenatbib]{mn2e}
\usepackage{graphicx}
\def\lesssim{\mathrel{\hbox{\rlap{\hbox{\lower4pt\hbox{$\sim$}}}\hbox{$<$}}}}
\def\gtrsim{\mathrel{\hbox{\rlap{\hbox{\lower4pt\hbox{$\sim$}}}\hbox{$>$}}}}
\def\msun{$M_{\odot}$}
\def\lteff{log\ $T_{\rm eff}$}

 \title[The born again (VLTP) scenario revisited]{The born again
 (VLTP)
scenario revisited: The mass of the remnants and implications for V4334 Sgr}
\author[M. M. Miller Bertolami and L. G. Althaus]{M. M. Miller
Bertolami$^{1,2}$\thanks{E-mail:
 mmiller@fcaglp.unlp.edu.ar} and
L. G. Althaus$^{1,2}$\thanks{E-mail: althaus@fcaglp.unlp.edu.ar}\\
$^{1}$Facultad de Ciencias Astron\'omicas y Geofisicas, UNLP, Paseo
 del
Bosque s/n, La Plata, B1900FWA, Argentina\\ $^{2}$Instituto de
Astrof\'{\i}sica La Plata, CONICET-UNLP, Paseo del Bosque s/n, La
 Plata,
B1900FWA, Argentina}
 
\begin{document}

\date{}

\pagerange{\pageref{firstpage}--\pageref{lastpage}} \pubyear{2007}

\maketitle

\label{firstpage}

\begin{abstract}

 We present 1-D numerical simulations of the very late thermal pulse
(VLTP) scenario for a wide range of remnant masses. We show that by
 taking
into account the different possible remnant masses, the
 observed evolution
of V4334 Sgr (a.k.a. Sakurai's Object) can be
 reproduced within the standard
1D-MLT stellar evolutionary models
 without the inclusion of any $ad-hoc$
reduced mixing efficiency. Our
 simulations hint at a consistent picture with
present observations of
 V4334 Sgr.  From energetics, and within the standard
MLT
 approach, we show that low mass remnants \hbox{($M\lesssim0.6$\msun )}
are
 expected to behave markedly different than higher mass remnants
\hbox{($M\gtrsim0.6$\msun)} in the sense that the latter are not expected to
expand significantly
 as a result of the violent H-burning that takes place
during the VLTP.  We
 also assess the discrepancy in the born again times
obtained by
 different authors by comparing the energy that can be liberated
by
 H-burning during the VLTP event.
\end{abstract}


\section{Introduction}

 Hydrogen(H)-deficient Post-Asymptotic Giant Branch (AGB) stars
display
 a wide variety of surface abundances, ranging from the
almost pure
 helium atmospheres of O(He) stars to the helium(He)-
carbon(C)- and
 oxygen(O)- rich surfaces of WR-CSPN and PG1159 stars
(see Werner \&
 Herwig 2006 for a review). In particular the surface
composition of
 the last group resembles the intershell region
chemistry of AGB star
 models when some overshooting in the pulse
driven convection zone
 (PDCZ) is allowed during the thermal pulses
(Herwig et al. 1997). For
 this reason, and due to the fact that the
occurrence of late
 (i.e. post-AGB) helium flashes is statistically
unavoidable in single
 stellar evolution modeling (Iben et al. 1983),
a late helium flash is
 the most accepted mechanism for the formation
of these stars (see,
 however, De Marco 2002). In particular during a
very late helium flash
 (VLTP; Herwig 2001b), as the H-burning shell
is almost extinguished,
 the PDCZ can reach the H-rich envelope.  As
a consequence H-rich
 material is carried into the hot C-rich
interior, and violently burned
 (see Miller Bertolami et al. 2006,
from now on Metal06, for a
 detailed description of the event). As
was already noted by Iben et
 al. (1983), the timescale in which H is
burned is similar to that of
 convective motions and consequently the
usually adopted instantaneous
 mixing approach is not valid. For this
reason only few numerical
 simulations of the VLTP exist in the
literature: Iben \& MacDonald
 (1995), Herwig et al. (1999), Herwig
(2001a), Lawlor \& MacDonald
 (2002), Lawlor \& MacDonald
 (2003)
and more recently Metal06.

The identification of V4334 Sgr (a.k.a. Sakurai's Object) as a star
undergoing
 a VLTP event (Duerbeck \& Benetti 1996) has renewed the
interest in this
 particular kind of late helium flash. V4334 Sgr has
shown a very fast
 evolution in the HR diagram of only a few years
(Duerbeck et
 al. 1997, Asplund 1999, Hajduk et al. 2005). In this
context it is worth mentioning that the
 theoretical born again times
(i.e. the time it takes the star to cross the HR
 diagram from a
white dwarf configuration to a giant star one) is a
 controversial
issue: whilst Iben \& MacDonald (1995), Lawlor \& MacDonald (2002) and
Metal06 obtain born
 again times of the order of one or two decades,
Herwig et al. (1999) obtain timescales of the order of centuries.

 The difference between the theoretical born again times of Herwig
et
 al. (1999) and the observed timescale of V4334 Sgr (and also V605
Aql,
 Duerbeck et al. 2002), prompted Herwig (2001a) to propose a
reduction in the
 mixing efficiency during the conditions of the
violent proton burning (by
 about a factor of 100 for their 0.604
\msun\ sequence) in order to match the
 observed timescale. However,
in view of the lack of hydrodynamical simulations
 of the violent
burning and mixing process during the VLTP, the idea has an
important drawback: it introduces a free parameter (i.e. the mixing
efficiency) that can only be calibrated with the situation that one
wants to
 study, thus losing its predictive power. Aside from this
philosophical aspect, current reduced mixing efficiency models (Herwig
2001a, Lawlor \& MacDonald 2003 and Hajduk et al. 2005) suffer from an
internal inconsistency. Indeed, contrary to what is stated in Herwig
(2001a) and Lawlor \& MacDonald (2003), changes in convective
velocities are expected to affect convective energy transport.  In
fact, as shown in appendix A, reducing convective velocities is
completely equivalent to reducing the mixing length. But more
importantly, the reproduction of the born again timescale does not
necessarily make models completely
 consistent with observations. In
fact, although Hajduk et al. (2005) claim
 that models with a
reduction in the mixing efficiency reproduce ``The
 real-time stellar
evolution of Sakurai's object'', a closer inspection shows
 that such
a claim should be taken with a pinch of salt. In particular the
effective temperature (and also the cooling rate) of the model in its
(first)
 return to the AGB contradicts the inferred effective
temperature (with 2
 different methods) during 1996-1998 (Duerbeck et
al. 1997, Asplund et
 al. 1999). In the same line, the extremely high
luminosity of their
 theoretical model in its first return to the AGB
($\sim 12\,500\,{\rm
 L}_\odot$ ) implies an extremely large distance
of more than 8 Kpc (by
 comparing with the values in Duerbeck et
al. 1997).  Surprisingly enough, this
 value is inconsistent with the
2 Kpc adopted by Hajduk et al. (2005)
 as well as with independent
distance estimations which place V4334 below 4.5
 Kpc and
preferentially between 1.5 and 3 Kpc (Kimeswenger 2002).
Hence, even if the reduction in the mixing efficiency by a factor 60
leads to
 born again times (Hajduk et al. 2005) similar to those
displayed by
 V4334 Sgr, that model fails to match other well
established observed
 properties. On the other hand, whilst the
reduced mixing efficiency models of Lawlor \& MacDonald (2003) do not
suffer from this inconsistencies, their models show high H-abundances
which are not consistent with observations of both V4334 Sgr or PG1159
type stars. 

In this context, it is worth noting that a strong reduction in the
mixing efficiency
 does not seem necessary in the VLTP models of Iben
\& MacDonald (1995) and
 Metal06 to reproduce observations. Most of
the existing computations of the VLTP have
 been performed for masses
in a very narrow range around the canonical mass
 $\sim0.6$ \msun. In
this context, we feel that an exploration of the (neglected)
importance of the mass of the remnant for the born again timescale
is
 needed. This is precisely the aim if this article. Indeed some
confusion seems to exist in the literature with respect to
 this
issue: whilst it is usually stated that low mass/luminosity models
evolve
 slower after a VLTP (Pollaco 1999, Kimeswenger 2002), it is
clear from
 Herwig's (2001a) 0.535 \msun\ model that lower
luminosities/masses lead to
 faster born again evolutions.
\section{Description of the work and numerical/physical details}
\begin{figure}
\begin{center}
  \includegraphics[clip, , width=8.4cm]{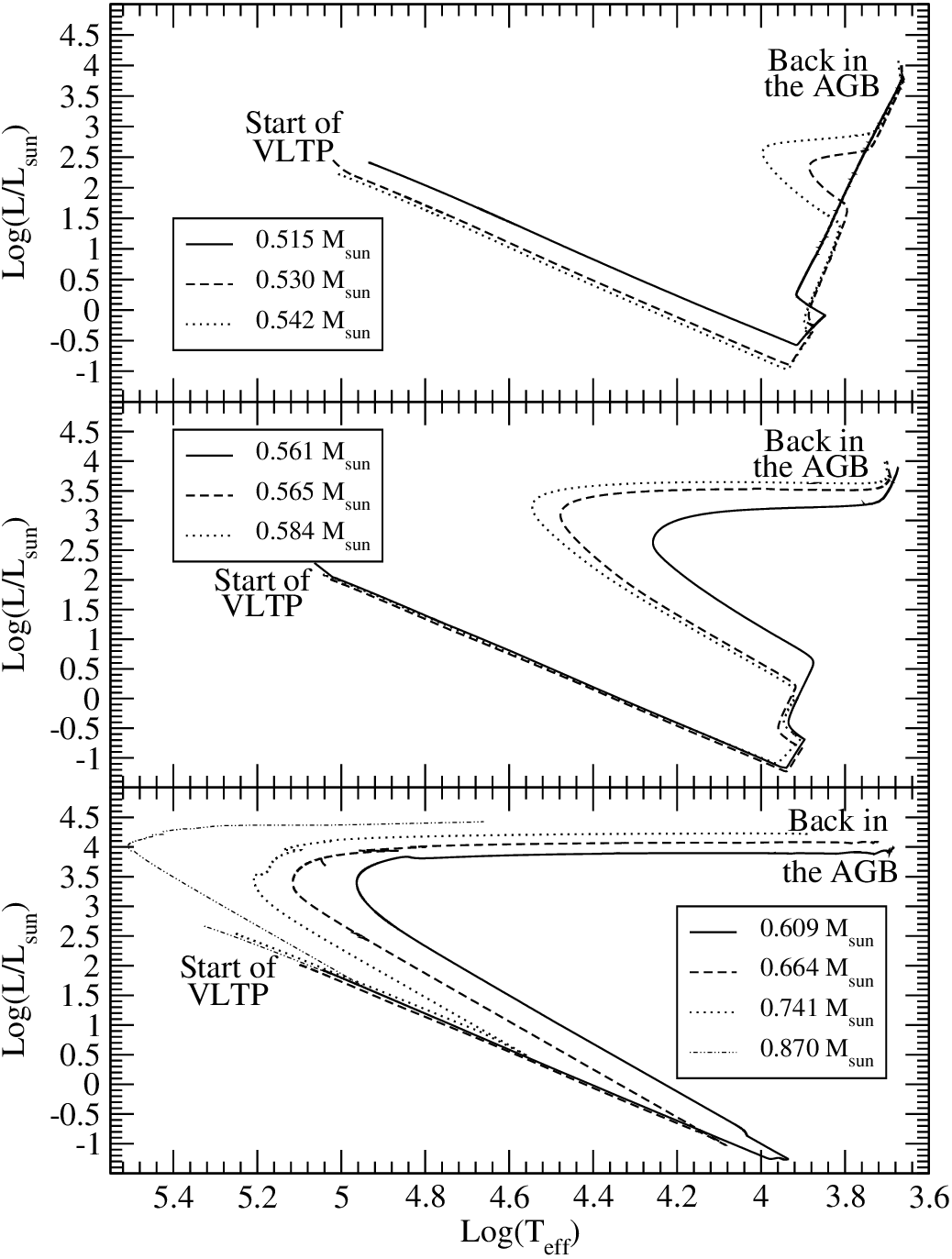} 
\caption{HR diagrams of the VLTP evolution for the sequences presented in this
  work during their first return to the AGB.}
\label{figura1} 
\end{center}
\end{figure}
\begin{table*}
\begin{center}
\begin{tabular}{|c|c|c|c|c|c|c|c|c|c|c|}\hline
Remnant  & Initial  & Ref.  & pre-flash & $t_{\rm BA}$ &  He-DCZ & H-DCZ & H-DCZ &  & & Loc. \\

  Mass  &  Mass     &   &   $M_{\rm H}$  &  &$\tau_{\rm loc}^1$-$\tau_{\rm loc}^2$ & $\tau_{\rm loc}^1$-$\tau_{\rm loc}^2$ &  $\tau_{\rm glo}$  & $\frac{^{12}{\rm C}}{^{13}{\rm C}}$& $\frac{^{12}{\rm C}}{^{14}{\rm N}}$ & H-burn peak    \\

 [\msun] & [\msun]  &            & [\msun]               & [yr] & [sec] &[sec]   & [sec]             &  &  & [\msun] \\\hline
 0.515   & 1       &   d        &  $2.24\times10^{-4}$   & 14   &  2183---2687  & 121---154 &  1931  & 6.6 & 13.4 & 0.486112 \\ 
 0.530   & 1        &   b        &  $2.30\times10^{-4}$   & 5   & 1399---1824& 30---39  & 326        & 6.  & 10.9 & 0.51202\\
 0.542   & 1       &   b        &  $1.75\times10^{-4}$   & 5.1   &  1260---1637& 33---42 & 333       & 6.3 & 12.1 & 0.52729 \\
 0.561   & 1.8     &   e        &  $1.00\times10^{-4}$   & 7.5   &  602---792 & 28---37 & 283        & 6.  & 11.2 & 0.54950  \\
 0.565   & 2.2     &   b        &  $8.19\times10^{-5}$   & 10.8   &  486---642& 28---35 & 329        & 6.1  & 11.1 & 0.55397 \\
 0.584   & 2.5     &   a        &  $8.70\times10^{-5}$   & 8.9   &  458---600  & 30---37 & 316         & 5.9 & 10.6 & 0.57485 \\
 0.609   & 3.05    &   b        &  $4.68\times10^{-5}$   & 157$^\dagger$   &  194---255 & 35---43 & 348  & 6.4 & 13.7 & 0.60166 \\
 0.664   & 3.5     &   b        &  $3.17\times10^{-5}$   & 106$^\dagger$   &  155---206 & 22---28 & 170  & 6.11 & 12.4 & 0.65968 \\
 0.741   & 3.75    &   c        &  $1.79\times10^{-5}$   & 65$^\dagger$   &  108---134  & 22---27 & 174  & 6.7 & 15.9 & 0.73822\\
 0.870   & 5.5     &   b        &  $8.85\times10^{-6}$   & ---$^*$   &  101---127  & 31---40 & 186       & 5.1$^*$ & 7.8$^*$ & 0.86885\\ \hline
\end{tabular}
\label{tabla}
\caption{Description of the sequences analysed in this work.  Fourth column
  shows the total amount of H ($M_{\rm H}$) left at the moment of the
  VLTP. Note the strong dependence of $M_{\rm H}$ on the mass of the
  remnant. $t_{\rm BA}$ stands for the time elapsed from the maximum of
  proton
 burning to the moment when the sequence reaches \lteff=3.8. Mean
  values of
 the local convective turnover timescales estimated with
  different
 prescriptions are shown for both He and H-driven convective
  zones (He-DCZ
 and H-DCZ respectively). The global estimation of the
  turnover timescale for
 the H-DCZ is also shown. Last column shows the
  approximate mass location of
 the maximum of proton burning at the moment
  of its maximum. $^\dagger$In
 these sequences the return to the AGB is
  mainly powered by the He-shell
 flash.$^*$ The 0.870 \msun\ sequence was
  stopped at high effective
 temperatures due to numerous convergence
  problems. References are: a-
 Metal06, b- Miller Bertolami \& Althaus
  (2006), c- C\'orsico et al. (2006),
 d- Althaus et al. (2007) and e-
  Unpublished.}
\end{center}
\end{table*}
In the present work we performed numerical simulations of the VLTP
scenario for several different remnant masses. The simulations have
been performed with the LPCODE (Althaus et al. 2005) by adopting the
Sugimoto (1970) scheme for the structure equations as described in the
Appendix A of Metal06. During the present work
we have adopted the standard MLT with a mixing length parameter
$\alpha=1.75$. Convective mixing was considered as a diffusive process
and solved simultaneously with nuclear burning as explained in Althaus
et al. (2003). We adopt a diffusion coefficient for the convective
mixing zones given by
\begin{equation}
  D=\frac{1}{3}\,l\, v_{\rm MLT}= \frac{\alpha^{4/3}\, H_P}{3} \left[\frac{c\,g}{\kappa\,\rho} (1-\beta) \nabla_{\rm ad} (\nabla_{\rm rad}-\nabla) \right]^{1/3}
\end{equation}
 as can be deduced from Cox \& Giuli (1968)\footnote{Note that there
 are two
typos in the expression for $D$ in footnote 6 of Metal06.}. As shown in
Metal06,
 no difference would arise if the expression of Langer et al. (1985)
was
 to be adopted. Diffusive overshooting was allowed at every convective
boundary with a value $f=0.016$ (see Herwig et al. 1997 for a
 definition of
$f$). By adopting the standard MLT we are ignoring the
 effect of chemical
gradients during the violent proton burning that
 can certainly influence the
results (Metal06). We
 choose not to include this effect for consistency, as
our overshooting
 prescription does not include the effect of chemical
gradients, and
 also for numerical simplicity. Also, as will be clear in the
following
 sections, we do not intend to reproduce the exact evolution of
born
 again stars (something that would require a much more sophisticated
treatment of convection) but instead to show the importance of the remnant
mass for the subsequent evolution.

 A detailed description of the
sequences considered in the present work is
 listed in Table 1.  HR diagram
evolution of the sequences during their first
 return to the AGB during the
VLTP is shown in Fig. 1. We mention that for more
 massive sequences this
will be their only return to the AGB as they only
 experience the He-driven
expansion. With exception of the sequence 0.561
 \msun, the prior evolution
of our VLTP sequences has been presented in our
 previous works (Metal06,
Miller Bertolami \& Althaus 2006, Corsico et al. 2006
 and Althaus et
al. 2007). All of them are the result of full and consistent
 evolutionary
calculations from the ZAMS to the post-AGB stage. In all the
 cases the
initial metallicity was taken as Z=0.02. One of the most
 remarkable features
displayed by Table 1 is the strong dependence of the total amount
 of H at
the moment of the VLTP ($M_{\rm H}$) on the mass of the
 remnant. This is
important in view of the discussion presented in Section
 4. Note however
that, aside from this strong dependence on the mass of the
 remnant,
$M_{\rm H}$ will also depend on the previous evolution and on the
 exact
moment at which the VLTP takes place. This is why the 0.584
 (0.530)\msun\
model has a higher $M_{\rm H}$ value than the 0.565
 (0.515)\msun\ model.

\section{Convection, timescales and time resolution}
As was already mentioned in Metal06, an extremely high time resolution
($\sim 10^{-5}$ yr) during the violent proton burning is needed in
order to avoid an underestimation of the energy liberated by proton
burning\footnote{In fact this can be one of the reasons for the
discrepancy in the born again times of different authors.}. Such small
time steps are close to the timescale needed to reach the steady state
described by the MLT (see Herwig 2001a). To be consistent with the
treatment of convection the time step should be kept above this value.

The convective turnover times can be estimated with different prescriptions
which can lead to significantly different values. In order to have a feeling
of what is really happening we have adopted three different estimations:
$\tau_{\rm loc}^1$, $\tau_{\rm loc}^2$  and $\tau_{\rm glo}$ (two local
and one global timescale), which are defined as:
\begin{equation}
\tau_{\rm loc}^1=\frac{1}{|N|}=\sqrt{\frac{kT}{\mu\,m_p\,g^2\,|\nabla-\nabla_{\rm ad}|}}
\end{equation}
\begin{equation}
\tau_{\rm loc}^2=\frac{\alpha\, H_P}{2\, v_{\rm MLT}}
\end{equation}
\begin{equation}
\tau_{\rm glo}=\int_{R_{\rm base}}^{R_{\rm top}} \frac{dr}{v_{\rm MLT}}
\end{equation}
The definition of the last two expressions is evident and the first is
the inverse of the Br\"unt-V\"ais\"al\"a frequency and provides the
timescale for the growth of convective velocities in a convectively
unstable region (see Hansen \& Kawaler 1994 for a derivation). In
Table 1 we list typical values of these timescales. Values at the
He-driven convection zone (He-DCZ) correspond to the moment just
before the violent proton ingestion (second stage of proton burning,
as described in Metal06) whilst values at the convective zone driven
by the violent proton burning (H-DCZ) correspond to the moment of
maximum energy release by proton burning (which is also the moment at
which important amounts of H start to be burned). For the local
estimations the averaged value over the whole convective zone is
displayed. It is worth mentioning that whilst global and local turn
over times coincide for the He-DCZ, the global estimation is, roughly,
one order of magnitude larger in the H-DCZ. From the values in Table 1
one would be tempted to state that by adopting a minimum allowed
timestep of $\sim 10^{-5}$ yr ($\sim315$s) at the beginning of the
violent stage of proton burning, we are allowing convective motions to
develop and, thus, being consistent with the steady state assumption
of the MLT.
However, a point should be mentioned. Hydrodynamical simulations of
``standard''
 (i.e. without proton ingestion) helium shell flashes
(Herwig et
 al. 2006) show that the steady state is not achieved in
only one turn
 over time. In fact Herwig et al. (2006) find that, at
a standard
 ($\epsilon_0=2\times10^{10}$ erg s$^{-1}$ gr$^{-1}$)
heating rate,
 about 10 turn over times are necessary. They also
find, however, that
 at an enhanced heating rate (30 times larger)
steady state is achieved
 about 2-3 times faster. Then, as the
heating rate at the base of the H-DCZ
 during the violent proton
burning is about
 $\epsilon=10^{13}$---$10^{14}$ erg s$^{-1}$
gr$^{-1}$ (1000---5000
 times larger than the one at the base of the
He-DCZ) one would expect
 that the steady state will probably be
reached in no much more than
 one turn over time. Consequently it
seems reasonable to choose a
 minimum allowed time step of $10^{-5}$
yr, which accordingly to our
 previous studies is needed to avoid an
underestimation of the energy
 liberated by proton
burning\footnote{Time steps of $\sim 10^{-5}$ yr
 are only needed
during the (short) runaway burning of protons where
 most of the H is
burned (see Metal06) and the energy liberated by
 proton burning is
$L_{\rm H}\sim 10^{7}- 10^{11} L_\odot$.}. We have also checked that
convective velocities remain subsonic during the stage of maximum
proton burning, being in all the cases $v_{\rm MLT}<0.12\, v_{\rm
sound}$. This is important because during the maximum of proton
burning convective velocities are high and the MLT is derived within
the assumption of subsonic convective motions. In our
 view the main
drawback of present simulations may come from the fact
 that
convective mixing differs from a diffusion process. In fact,
 during
a VLTP, H needs to be mixed with $^{12}$C in order to be
 burned. It
is probable that, initially, H may be transported downwards
 in the
form of plumes and thus no important amounts of H would be
(initially) mixed with $^{12}$C.  This situation would be very
different from the picture of diffusive mixing with or without
reduced
 mixing efficiency. How far our results are from reality will
depend on
 how far the place at which most H is burned differs from
the one
 predicted by diffusive mixing within the standard (without
reduced
 mixing efficiency) MLT.

\section{Energetics}
\begin{figure}
\begin{center}
  \includegraphics[clip, , width=8.4cm]{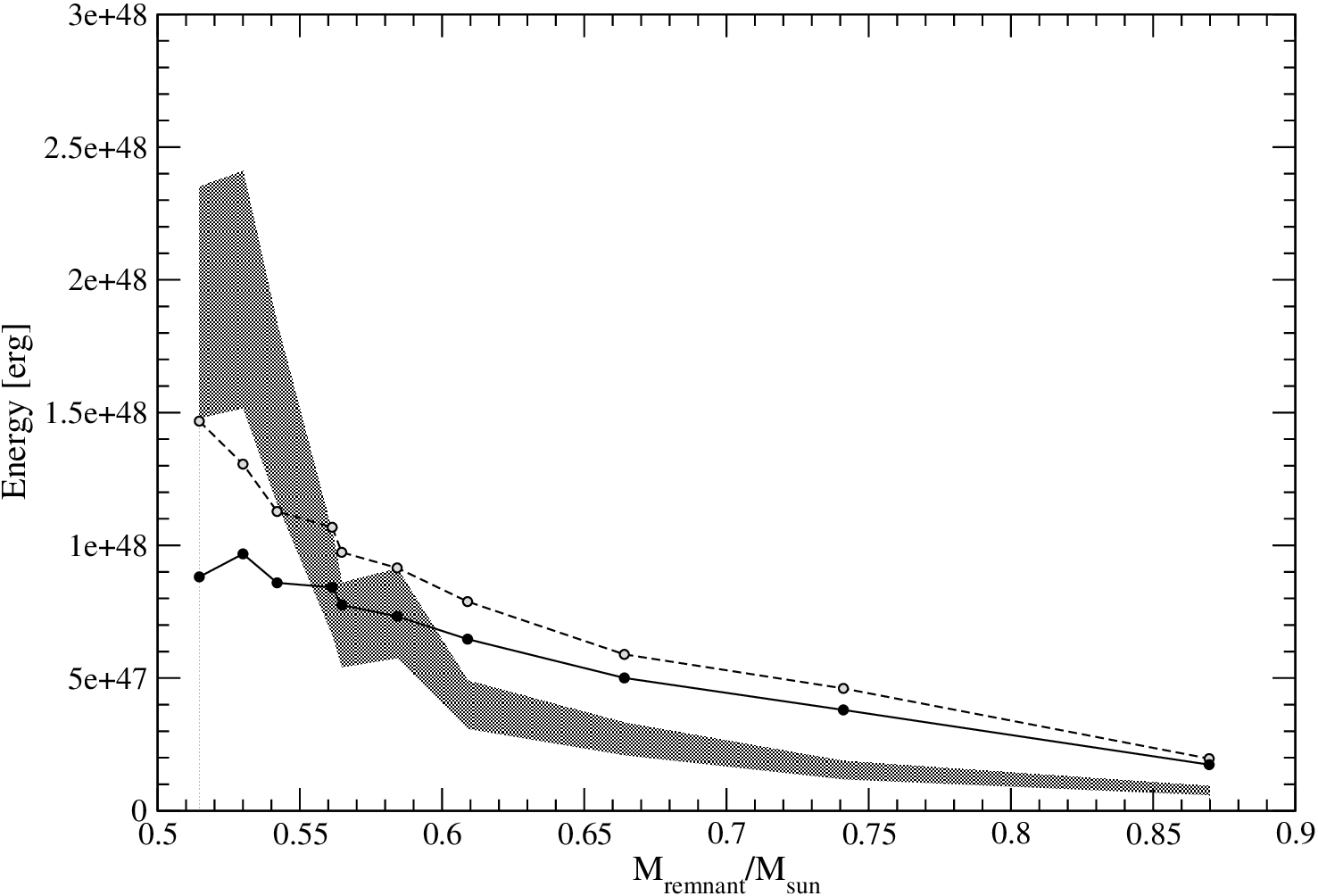} 
\caption{Comparison of the total amount of energy that can be liberated during
the violent proton burning ($E_{\rm H}$; shaded region) with the energy
necessary to expand the envelope above the point of maximum proton burning
($|E_{\rm tot}|$; solid line). Broken line shows the gravitational binding
energy of that envelope ($|E_{\rm g}|$).}
\label{figura} 
\end{center}
\end{figure}
As mentioned in the introduction, born again times from different
authors can differ by almost two orders of magnitude. Also, as found
in Metal06 for $\sim0.59$ \msun\ models, numerical issues can change
the born again times completely and even suppress the H-driven
expansion (if energy from H-burning is underestimated by a factor
$\sim2$). Regarding the occurrence or not of the H-driven expansion it
is also interesting to note the sudden change in the born again times
that can be seen in Table 1 between the 0.584 and the 0.609 \msun\
models. Whilst the 0.584 \msun\ model shows a H-driven expansion and
consequently short born again times (8.9 yr), the 0.609 \msun\ model
does not, displaying born again times of 157 yr.  In this context we
feel it interesting to analyse the energetics of the VLTP. To this
end, we estimate the energy that can be liberated by proton burning if
the whole H-content of the star is burned ($E_{\rm H}$). This is shown
in Fig. 2 as a shaded zone. This estimation can be done by noting that
during the VLTP (almost) all the H is burned first through the chain
$^{12}$C+p$\rightarrow$$^{13}$N+$\gamma\rightarrow$$^{13}$C+e$^+$+$\nu_e$
(which liberates 3.4573 MeV per proton burned) and then through
$^{12}$C+p$\rightarrow$$^{13}$N+$\gamma\rightarrow$$^{13}$C+e$^+$+$\nu_e$
and $^{13}$C+p$\rightarrow$$^{14}$N+$\gamma$ working at the same rate
(a process that liberates 5.504 MeV \emph{per proton} burned,
Metal06). These two values give rise to the lower and upper boundary
in Fig. 2. We also calculate the gravitational binding energy $E_{\rm
g}$, just before the violent stage of proton burning, of the zone
above the peak of proton burning. From numerical models we know this
is the zone that expands due to the energy liberated by the H-burning
(and will be denoted as \emph{envelope} in the following)
\begin{equation}
E_{\rm g}=-\int_{\rm envelope} \frac{Gm}{r} dm 
\end{equation}
We also estimate the internal energy of the envelope $E_{\rm i}$ as 
\begin{equation}
E_{\rm i}=\int_{\rm envelope} T c_v dm
\end{equation}
 From these values we compute the total energy of the envelope
$E_{\rm tot}$ (which is negative for a gravitationally bound system) as
\begin{equation}
E_{\rm tot}=E_{\rm g}+E_{\rm i}
\end{equation}
$|E_{\rm tot}|$ is then the energy needed to expand the envelope to
 infinity. Its value is lower than $|E_{\rm g}|$ because (the Virial
 theorem enforces) as the envelope expands it cools and internal
 energy is released, helping the expansion. By comparing $E_{\rm H}$
 and $|E_{\rm tot}|$ we can resonably decide if the burning of H in a
 given
 remnant can drive an expansion of its envelope. The result
 is
 surprising, as inferred from Fig. 2. Note that for models above
 0.6
 \msun\ the energy that can be liberated by proton burning
 \emph{is
 not enough} to expand the envelope. This result is
 completely
 consistent with what is displayed by our detailed
 modeling of the
 process. Although this extreme consistency with
 such rough estimation
 can be just a coincidence, the different
 behaviour of $E_{\rm tot}$
 and $|E_{\rm H}|$ with the mass of the
 remnant is worth
 emphasising. Whilst the former decreases only
 slightly with the stellar mass, the later shows a steeper
 behaviour. This is expected
 because the total H-amount of a remnant
 is a steep function of the
 stellar mass. This fact yields a
 transition at a certain stellar mass value below which the H-driven
 expansion is possible and above which the H content of the remnant is
 not enough for
 H-burning to trigger an expansion.
 
 This
 energetic point of view not only holds the clue to understand the
 distinct behaviour of our sequences but also helps to understand
 the
 differences in previous works. Note that models with masses
 close to the $\sim
 0.6$ \msun\ transition value will be very
 sensitive to numerical issues that
 can eventually alter the release
 of H-burning energy. This could explain why
 an underestimation by
 factor 2 in the H-burning energy reported in Metal06
 strongly
 affects the born again times.  It also helps to explain why Althaus
 et al. (2005), in which no extreme care of the time step during the
 violent
 proton burning was taken, reported longer born again times
 (20-40 yr) than Metal06
 for the same initial model\footnote{This
 was also due to a difference in the
 definition of the diffusion
 coefficient by a factor 3 (see Metal06).}. In this view, the short
 born again times found by Lawlor \& MacDonald (2002) are probably due
 to the low mass of their models (0.56-0.61 \msun). Also the
 difference in born again times between Iben \& MacDonald (1995) and
 Herwig et
 al. (1999) simulations can probably be understood in this
 context. In fact, whilst Iben \&
 MacDonald (1995) and Herwig et
 al. (1999) sequences have similar remnant
 masses (0.6 \msun\ and
 0.604 \msun, respectively) the total amount of H is markedly
 different in both cases, being $\sim 2.37 \times 10^{-4}$ \msun\ and
 $\sim 5. \times
 10^{-5}$ \msun , respectively. This leads to
 energies ($E_{\rm H}$) of
 3.29---5.23 $\times10^{47}$ erg for
 Herwig et al. (1999) (similar to the
 values of our own 0.609 \msun\
 sequence, see Fig. 2) and 1.56---2.48
 $\times10^{48}$ erg for Iben
 \& MacDonald (1995). This difference of almost a
 factor of five in
 the energy released by proton burning is very probably
 the reason
 why Iben \& MacDonald (1995) find a H-driven expansion whilst
 Herwig et al. (1999) do not.

 On the basis of these arguments, the finding of Herwig (2001a) about
 that a reduction in the mixing efficiency can lead to shorter born
 again times can be understood as follows. By reducing the mixing
 efficiency the point at which the energy is liberated by proton
 burning is moved outwards. Then the value of $|E_{\rm tot}|$ is
 lowered whilst the total amount of H remains the same and thus no
 change in $E_{\rm H}$ will happen. Then, changing the mixing
 efficiency moves the solid line in Fig. 2 up and down, which alters
 the transition mass value at which H-driven expansion begins to be
 possible\footnote{There is also a secondary effect which comes from
 the fact that the closer to the surface the energy is released, the
 shorter it takes the liberated energy to reach the surface of the
 star.}.
 
\section{Comparison with the observed behaviour of V4334 Sgr}
%
%
%
\begin{figure}
\begin{center}
  \includegraphics[clip, , width=8.4cm]{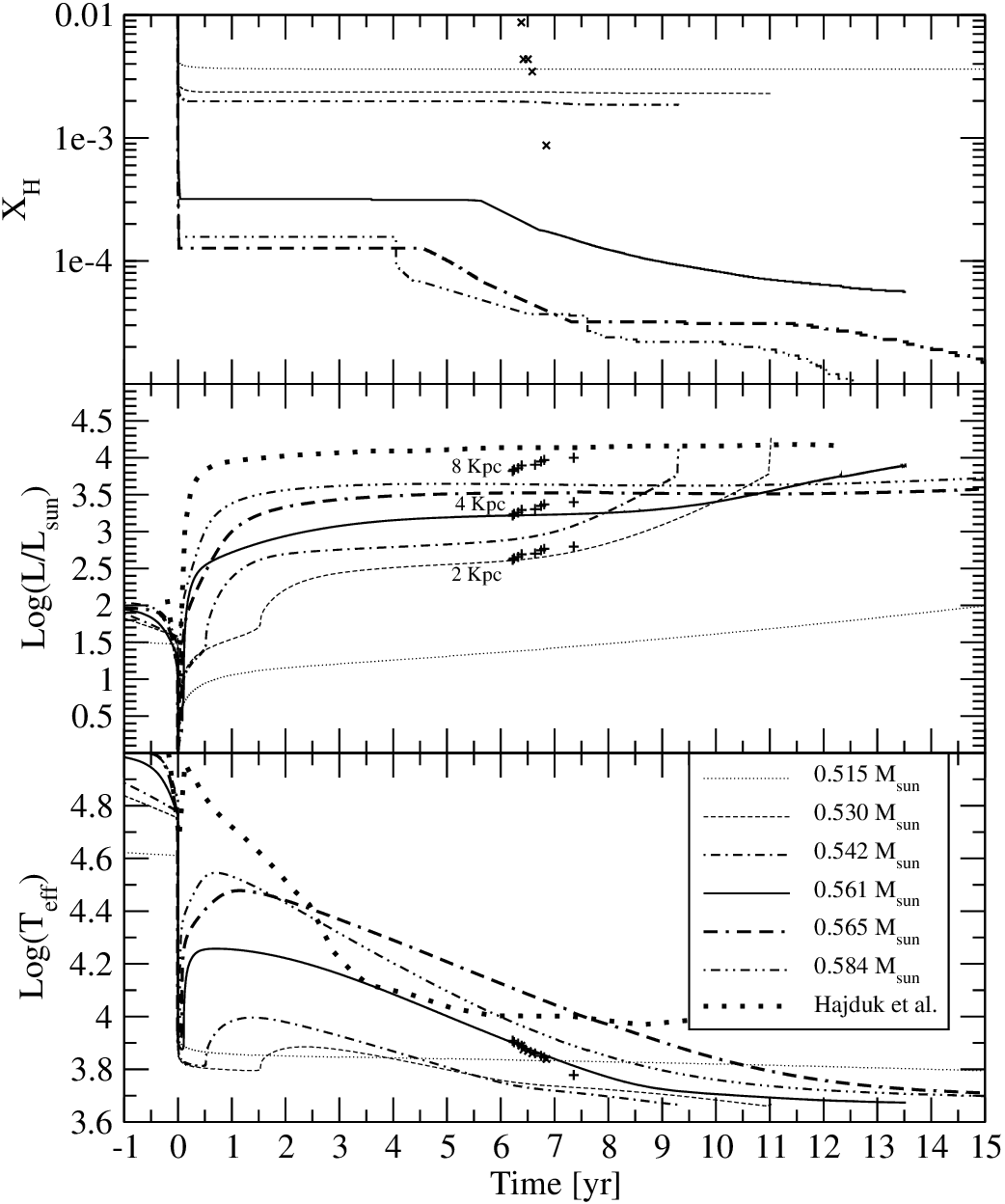} 
  \caption{Bottom and middle panel show the evolution of luminosity
 and
  effective temperature of the models after the VLTP (set at 0 yr)
 compared
  with the observations of V4334 Sgr (Duerbeck et al. 1997,
 Asplund 1999; +
  and $\times$ signs respectively). The evolution of
 luminosity and
  effective temperature of Hajduk et al. (2005),
 extracted from Fig. 2
  of that work, is shown for comparison. The zero point in the x-axis of the
  observations was
 arbitrarily set to allow comparison with the
  models. Upper pannel
 shows the evolution of the H abundance at the
  outermost layer of the
 models (which should be close to the surface
  abundance) compared
 with the observed abundances at V4334 Sgr (Asplund
  1999).}
\label{figura3} 
\end{center}
\end{figure}
\begin{figure}
\begin{center}
  \includegraphics[clip, , width=8.4cm]{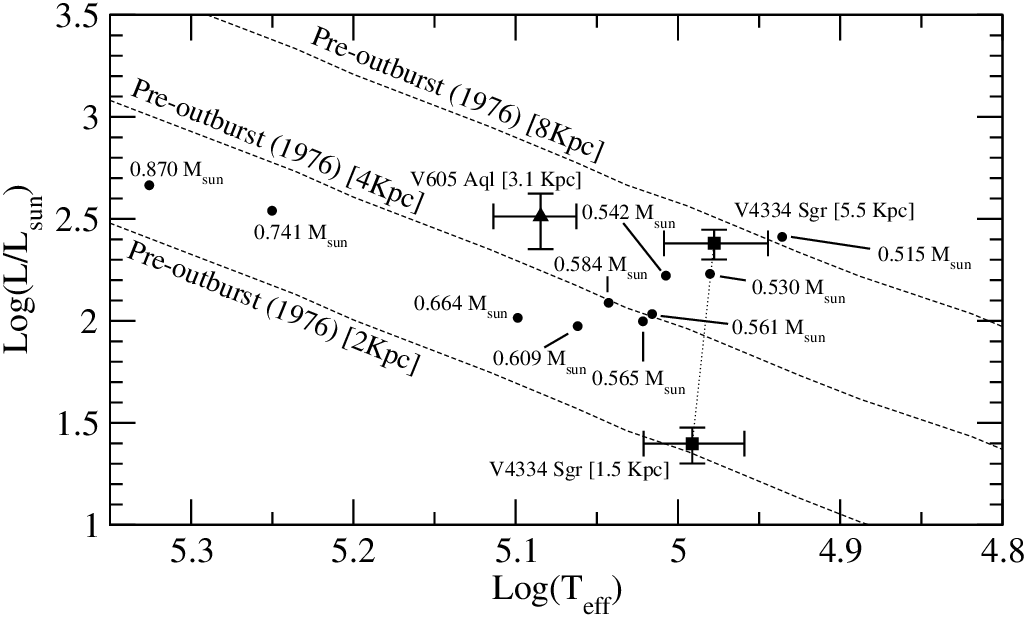} 
  \caption{Location of the models in the HR diagram at the moment
  of the VLTP. The lines show the possible 1976 detection of V4334
  (taken from Herwig 2001a and rederived for different assumed
  distances). Also inferred pre-flash location of V4334 Sgr and V605
  Aql (Kerber et al. 1999, Lechner \& Kimeswenger 2004) are shown for
  comparison. Note the strong dependence of the $T_{\rm eff}$ on the mass of
  the remnant.}
\label{figura4} 
\end{center}
\end{figure}
In Fig. 3 we compare the evolution of luminosity and effective
 temperature
of our sequences with that observed at V4334 Sgr (Duerbeck
 et al. 1997,
Asplund et al. 1999). Also the preoutburst location of
 V4334 Sgr and V605
Aql (from Kerber et al. 1999 and Herwig 2001a) is
 compared with the location
of the sequences at the VLTP in
 Fig. 4. Luminosities have been rederived for
different distances to
 allow comparison. This was done by making use of the
fact that
 interstellar extintion is supposed to be constant in that
direction of
 the sky for distances above 2 Kpc. For remnants that display
a
 H-driven expansion ($\lesssim 0.584$ \msun) the main feature can be
described as follows. Whilst the preoutburst location of least massive
remnants ($\lesssim 0.542$\msun) is compatible with higher distances
($>4$Kpc) their luminosity curves are compatible with lower distances
($<4$Kpc). However, for the 0.561, 0.565 and 0.584 \msun\ models, 
preoutburst locations are compatible with a distance of $\sim 4$ Kpc
 in
agreement with the distance inferred from their luminosity curves (4-6 Kpc).
Its is worth noting that all models between 0.530 and 0.584
 \msun\ display
cooling rates which are compatible with that observed
 in V4334 Sgr. Also,
their born again times range from 5 to 11 yr not
 far from that observed at
V4334 Sgr (or V605 Aql, Duerbeck et
 al. 2002).
 
 It is particularly
interesting to note the behaviour of our 0.561 \msun\
 model. It reaches the
effective temperature at which V4334 Sgr was discovered
 in about $\sim6$ yr
after the VLTP, showing at that point a very similar
 cooling rate as that
measured at V4334 Sgr with two different methods
 (Duerbeck et al. 1997,
Asplund et al. 1999). On the other hand, its luminosity
 at the moment of
discovery would imply a distance of about 3-4 Kpc which is
 completely
consistent with preoutburst determinations when that distance is
 adopted.

 As shown in Table 1 all of our models predict $^{12}$C/$^{13}$C$\sim
 6$
(by mass fractions) not far from the inferred $^{12}$C/$^{13}$C at V4334 Sgr
(Asplund
 et al. 1999). Our models also predict high $^{14}$N abundances at
their
 first return to the AGB (see Table 1), at variance with Herwig (2001b)
but in
 agreement with observations of V4334 Sgr (Asplund et al. 1999).

In Fig. 3 the evolution of the reduced mixing efficiency model of Hajduk et
al. (2005) is shown for comparison. Although this model evolves initially
faster than our sequences the behaviour of its luminosity and temperature
are
 not consistent with those observed at V4334 Sgr. In particular all the
effective temperature determinations of V4334 Sgr in its first return to the
AGB lie beyond
 the minimum effective temperature attained by that
sequence. Also its
 luminosity at low effective temperatures implies a
distance of d$>8$ Kpc, in
 contradiction with all independent distance
determinations (Kimeswenger 2002).

 As it is shown in the top pannel of Fig. 3, the 0.561, 0.565 and
0.584
 \msun\ sequences ---those which are both compatible with pre-
and
 post- outburst inferences for the temperature and luminosity of
V4334
 Sgr --- also reproduce qualitatively the drop at low
effective
 temperature in the H abundance of V4334 Sgr observed by
Asplund et
 al. (1999). Note however that there is a quantitative
disagreement by
 more than one order of magnitude between
observations and model
 prediction. This may be either due to an
intrinsic failure of the
 models or because the quantities plotted
are not exactly the surface
 abundances of the models but, instead,
the H-abundance at the
 outermost shell of the models (which can
differ from the actual surface value).

As was already noted by Metal06,
models without reduced mixing
 efficiency fail to reproduce the fast
reheating of V4334 Sgr (as
 reported by Hajduk et al. 2005). One may argue
that this can be due to
 the fact that mass loss was ignored in the models,
whilst V4334 Sgr
 displayed strong mass loss episodes once its temperature
dropped below
 $\sim 6000$K. In fact if we impose a mass loss rate similar to
that inferred
 in V4334 Sgr ($2\times10^{-4}$ \msun/yr, Hajduk et al. 2005) we
find that
 the 0.561 \msun\ model reheats (reaches temperatures greater
than
 10000 K) in only 27 yr and the 0.584 \msun\ model does it in 24
yr. This is certainly faster than in the absence of mass loss but
 still a
factor of $\sim 3-4$ greater than what was observed at V4334
 Sgr. However,
this failure to reproduce the late photospheric
 evolution of V4334 Sgr
should not be a surprise if we consider the
 fact that the hydrostatic
equilibrium condition is completely broken
 in the outer layers once the born
again star reaches the low
 temperature/high luminosity region of the HR
diagram, and thus there
 is no reason to expect that present hydrostatic
sequences will
 reproduce reality accurately at that point of 
evolution.

In any case, the present simulations show that it is
possible to
 attain short born again times without imposing an $ad-hoc$
reduction
 of the mixing efficiency if different remnant masses are
allowed. Even
 more, this approach (aside from not introducing a free
parameter) leads
 to models which are more consistent with the observations
of V4334 Sgr
 than models with a reduced mixing efficiency (Herwig 2001a,
Hajduk et
 al. 2005).

%
%
%
%

%
%
\section{Discussion and final remarks}
In the present work we have presented 1-D hydrostatic evolutionary
 sequences
of the VLTP scenario for different remnant masses. In
 Section 4 we have made
an analysis of the energetics of the VLTP that
 shows the importance of the
mass of the remnant for the born again
 timescale. In particular, that
argument shows that, within the standard
 MLT with no reduction of the mixing
efficiency, it is expected that the H-driven expansion that leads to short
born again times will only be present in VLTPs of low mass
 remnants. The
transition remnant mass value below which the H-driven expansion
 (i.e. a
short born again timescale) takes place is close to
 the canonical mass value
$\sim 0.6$\msun. The precise value of this transition mass depends on the
exact location at which most H is
 burned, and thus on a detailed description
of the mixing and burning
 process. A more accurate value of the transition
mass will have to wait
 until hydrodynamical simulations of the violent
H-ingestion became
 available. We have also shown that the energetic point of
view discussed in Section 4
 can help to understand the differences in the
calculated born again
 times by different authors.

We have also compared our predictions with the observations of real VLTP
objects (mainly
 V4334 Sgr.). As was noted early in this work we do not
expect from
 such simplified models and treatment of convection (as those
presented in this
 and all previous works) to reproduce the exact evolution
observed at real VLTP
 stars (like V4334 Sgr or V605 Aql). In fact this is
one of the reasons why
 inferences such as the need of a reduction in the
mixing efficiency coming
 from the fitting of the born again times of 1D
hydrostatic sequences for a
 single remnant mass should be taken with
care. Also the born again times given
 in Table 1 should be taken with care,
as for example a reduction by a factor
 of three in the mixing efficiencies
leads to a reduction of a factor of two in the
 born again times of one of
the sequences presented by Metal06. However, as
 argued in Section 2, we have
some reasons to believe that present models may
 not be that far from
reality. In this context the comparison with observations in Section 5 shows
that it is possible to roughly reproduce the observed behaviour of
 V4334 if
a mass of $\sim 0.56$ \msun\ for the remnant is
 assumed and a distance of
$\sim$ 3-4 Kpc is adopted. Interestingly enough,
 these distances are similar
to the ones derived by most distance
 determination methods (Kimeswenger
2002). What makes this model very
 interesting is not only that it avoids the
inclusion of a completely
 free parameter but that it solves the
inconsistency problems of
 reduced efficiency models (Herwig 2001a, Hajduk et
al. 2005) mentioned
 in the Introduction (and that can be appreciated in
Fig. 3). Interestingly enough, $\sim 0.56-0.59$ \msun\ sequences also
reproduce qualitatively the drop in the H abundance observed in V4334
 Sgr at
low effective temperatures. As this late drop in H-abundance is
 partially
due to dilution of the outer layers of the envelope into
 deeper layers of
the star (and also due to a deepening of the shell at
 which $\tau_{2/3}$ is
located), it is expectable that this drop in H
 would be accompanied by a
raise in s-process elements abundances, just
 as observed in V4334 Sgr.
 
The main drawback of our models is that they fail to reproduce the
 fast
reheating of V4334 Sgr (Hajduk et al. 2005) by about a factor of
 4. However
as discussed in Section 5 this is not a surprise as one of
 the main
hypothesis of the modeling (that of hydrostatic equilibrium)
 is explicitly
broken in the outer layers of the models once the VLTP
 star is back to the
AGB at very low tempetures. Thus, there is no
 reason to expect that the
sequences of this work will accurately
 reproduce reality at that course of
 evolution.
 
 We conclude then, that V4334 Sgr is not
``incomprehensible'' (Herwig
 2001a) within the standard MLT approach and
that there is ``apriori'' no need for a reduction of the mixing
efficiency. Needless to
 say, this statement does not imply that mixing
efficiency is not
 reduced during the proton ingestion in the VLTP nor that
the MLT
 approach is correct during the VLTP ---something that will only be
known once hydrodynamical simulations of the H ingestion and burning
 become
available---, but only shows that it is possible to roughly
 reproduce the
observations within the standard MLT approach.
 
 In any case, the main
conclusion of the present work is that different remnant
 masses \emph{have
to} be considered when comparing theoretical expectations
 with real VLTP
objects.

\section*{Acknowledgments}

 This research was supported by the Instituto
 de Astrof\'{\i}sica
La Plata and by PIP 6521 grant from CONICET. M3B
 wants to thank the
Max Planck Institut f\"ur Astrophysik in Garching
 and the European
Assocciation for Research in Astronomy for an
 EARA-EST fellowship
during which the central part of this work was
 conceived. We warmly
thank A. Serenelli, K. Werner, M. Asplund and an anonymous referee
for
 a careful reading of the manuscript and also for comments and
suggestions
 which have improved the final version of the
article. M3B wants to thank A. Weiss for useful discussions about
convection. We also thank
 H. Viturro and R. Martinez for technical
support.

\appendix

\section{Consistent treatment of reduced convective velocities.}
It has been proposed in some articles (Herwig 2001a, Schlattl et
al. 2001) that convective mixing efficiency can be reduced during 
violent H-flashes. Even more, Herwig (2001a) mentions that convective
material transport can be changed in main sequence stellar models by
orders of magnitude without any change in stellar parameters. Going
even further Lawlor \& MacDonald (2003) conclude that, as reduced
mixing efficiency does not produce significant changes in usual
stellar evolutionary stages, reduced mixing velocities could be usual
in stellar evolution. In what follows it is shown that, if a reduction
of mixing velocities is considered in the full treatment of the MLT,
the effect of changing the convective velocities is undistinguishable
from changing the value of $\alpha$ (the mixing length to pressure scale
height ratio). Then the previous conclusions come from an
inconsistent treatment of the reduction in mixing velocities.
\subsection{Reduced convective velocites and convective energy flux}
Following Kippenhahn \& Weigert (1990) we can derive the mean work
done by buoyancy on the convective elements to be:
\begin{equation}
\bar{W(r)}=g\delta(\nabla-\nabla_e)\frac{{l_m}^2}{8 H_P},
\end{equation}
where the symbols have their usual meanings
Then to propose a reduced mixing velocity within the MLT, we propose
that only a fraction of the work done by buoyancy goes into the
kinetic energy of the convective elements. We can write this as:
\begin{equation}
\bar{v}=f_v \left(\frac{g\delta(\nabla-\nabla_e)}{8 H_P}\right)^{1/2}l_m.
\end{equation}
Here $f_v$ is a new (and, a priori, ``free'') parameter giving the
factor by which the standard value of $\bar{v}$ is reduced.  Following
Kippenhahn \& Weigert (1990) and inserting this relation in the equation for
convective energy transport,
\begin{equation}
F_{\rm con}= \rho v c_P DT,
\end{equation}
and replacing $DT$ in terms of $(\nabla-\nabla_e)$ we get:
\begin{equation}
F_{\rm con}=\frac{C_P \rho T \sqrt{g \delta}}{4\sqrt{2}}{l_m}^2 \left(\frac{(\nabla-\nabla_e)}{H_P}\right)^{3/2}f_v.
\end{equation}
The convective flux is thus accordingly reduced by the factor $f_v$.
\subsection{Modified dimensionless equations}
From the relation
\begin{equation}
\left[\frac{\nabla_e-\nabla_{\rm ad}}{\nabla-\nabla_e}\right]=\frac{6acT^3}{\kappa \rho^2 C_P l_m \bar{v}}
\end{equation}
and defining the usual dimensionless quantities $U$ and $W$:
\begin{equation}
U=\frac{3acT^3}{c_P \rho^2 \kappa {l_m}^2}\sqrt{\frac{8H_P}{g\delta} }
\end{equation}
\begin{equation}
W=\nabla_{\rm rad}-\nabla_{\rm ad}
\end{equation}
we get new equations (equivalent to equations 7.14 and 7.15 in
Kippenhahn \& Weigert 1990) for $U$:
\begin{equation}
(\nabla_e-\nabla_{\rm ad})=2\frac{U}{f_v}\sqrt{\nabla-\nabla_e}
\end{equation}
\begin{equation}
(\nabla-\nabla_e)^{3/2}=\frac{8}{9}\frac{U}{f_v} (\nabla_{\rm rad}-\nabla)
\end{equation}
It seems natural then to define the quantity $U'=\frac{U}{f_v}$. 
\begin{equation}
U'=\frac{3acT^3}{c_P \rho^2 \kappa {l_m}^2 f_v}\sqrt{\frac{8H_P}{g\delta} }
\end{equation}
Then
defining the quantity $\zeta=+\sqrt{\nabla-\nabla_{\rm ad}+U'^2}$ we
get that the new dimensionless cubic equation for the MLT is:
\begin{equation}
(\zeta-U')^3+\frac{8U'}{9}\left(\zeta^2-U'^2-W \right)=0
\end{equation}
This becomes the new equation to be solved in order to obtain the real
value of $\nabla$. This is the same equation than the 7.18 one given in
Kippenhahn \& Weigert (1990), being the only difference the definition
of $U'$.
\subsection{The diffusion coefficient}
In the context of diffusive convective mixing, material transport and
mixing is ruled by the diffusion coefficient, which is usually defined
as $D=\frac{1}{3}l_mv$. Taking Eq. A2, replacing
$(\nabla-\nabla_e)$ with Eq. A9 and using the definiton of $U'$ we
get for the mean velocity of convective motions the expression:
\begin{equation}
 \bar{v}=\left(\frac{{f_v}^2 l_m}{H_P}\right)^{1/3}\left[\frac{cg\nabla_{\rm ad}(1-\beta)}{\rho \kappa}(\nabla_{\rm rad}-\nabla) \right]^{1/3}
\end{equation}
Introducing this into the definition of $D$ we get (using $l_m=H_P \alpha$):
\begin{equation}
D=\frac{1}{3}(\alpha^2f_v)^{2/3}H_P\left[\frac{cg(1-\beta)}{\rho \kappa} \nabla_{\rm ad} (\nabla_{\rm rad}-\nabla) \right]^{1/3}
\end{equation}
\subsection{Conclusion}
As $U'$ (and thus $\nabla$) and $D$ depend only on the product
$\alpha^2 f_v$ (and not on both quantities independently) then changes
in the mixing length and changes in the mixing velocities are
indistinguishable, making it unnecessary to consider a new free
parameter within the MLT. A reduction in the mixing efficiency should
be regarded as equivalent to reducing the mixing length. Then, as a
consequence, changing the mean velocity of convective motions would
produce an appreciable difference in any case in which non-adiabatic
convection takes place (lower main sequence, RGB, AGB stellar models).

\end{document}